\documentclass[aps,pra,twocolumn,groupedaddress,10pt,showpacs]{revtex4-1}
\usepackage{graphicx}
\usepackage{amsfonts}
\usepackage{amsmath}
\usepackage{amssymb}
\usepackage{epsfig}
\usepackage{color}
\usepackage{hyperref}
\usepackage{float}
\usepackage{breqn}

\newcommand{\ket}[1]{| #1\rangle}

\DeclareMathOperator\erfc{erfc}

\newcommand{\uchicago}{Department of Physics and James Franck Institute, University of Chicago, 929 E. 57$^{th}$ St., Chicago, IL 60637, USA}

\begin{document}
\title{Photons and polaritons in a time-reversal-broken non-planar resonator}
\author{Jia Ningyuan}
\author{Nathan Schine}
\author{Alexandros Georgakopoulos}
\author{Albert Ryou}\thanks{Present Address: Department of Electrical Engineering, University of Washington, 185 Stevens Way, Seattle, WA 98195, USA.}
\author{Ariel Sommer}\thanks{Present Address: Department of Physics, Lehigh University, 16 Memorial Drive East, Bethlehem, PA 18015, USA.}
\author{Jonathan Simon}
\affiliation{\uchicago}
\date{\today}

\begin{abstract}
From generation of backscatter-free transmission lines, to optical isolators, to chiral Hamiltonian dynamics, breaking time-reversal symmetry is a key tool for development of next-generation photonic devices and materials. Of particular importance is the development of time-reversal-broken devices in the low-loss regime, where they can be harnessed for quantum materials and information processors. In this work, we experimentally demonstrate the isolation of a single, time-reversal broken running-wave mode of a moderate-finesse optical resonator. Non-planarity of the optical path produces a round-trip geometrical (Pancharatnam) polarization rotation, breaking the inversion symmetry of the photonic modes. The residual time-reversal symmetry between forward-$\sigma^+$/ backwards-$\sigma^-$ modes is broken through an atomic Faraday rotation induced by an optically pumped ensemble of $^{87}$Rb atoms residing in the resonator. We observe a splitting of 6.3 linewidths between time-reversal partners and a corresponding optical isolation of $\sim$ 20.1(4) dB, with 83(1)\% relative forward cavity transmission. Finally, we explore the impact of twisted resonators on T-breaking of intra-cavity Rydberg polaritons, a crucial ingredient of photonic materials and specifically topological optical matter. As a highly coherent approach to time-reversal breaking, this work will find immediate application in creation of photonic materials and also in switchable narrow-band optical isolators.
\end{abstract}

\maketitle

Within the condensed matter community there is a growing interest in creating synthetic material analogs made of light to explore idealized models which are difficult to realize within the solid state. In such ``photonic materials,'' photons in either the optical- or microwave- domain may be made to behave as massive particles that are trapped and allowed to interact with one another. Using arrays of micro-fabricated waveguides~\cite{rech2013phot} and resonators~\cite{hafe2013imag,ning2015time}, or exotic Fabry P\'erot cavities~\cite{schine2016synthetic, klae2010bose, sommer2016engineering}, it has even become possible to engineer the single-particle photonic dispersion to create gauge fields for these massive photons. To mediate interactions between photons they must be coupled to matter--to Josephson junctions in the microwave domain~\cite{wallraff2004strong,houck2012chip}, and either to Rydberg-dressed atoms ~\cite{peyr2012quan, dudi2012stro, pari2012obse, firs2013attr, Jia2016CavityRydPol, jia2017strongly} or other nonlinear emitters ~\cite{sun2016quantum} in the optical domain. A crucial missing ingredient is the ability to explicitly break time reversal symmetry without spoiling the exquisite longevity of the photonic particles. In the ring resonators or waveguides described above, such time-reversal symmetry breaking would energetically preclude backscattering, which would otherwise correspond to reversal of synthetic gauge fields, and more broadly to physics beyond the material dynamics under consideration. In interacting systems enforcing such a T-broken single particle sector is more crucial, as the interactions themselves will otherwise violate the symmetry which protects the topological character of the system~\cite{fialko2014fragility,lodahl2016chiral}.

In the optical domain, time-reversal breaking has long been employed in isolators, where the Faraday effect provides a non-reciprocal polarization rotation. However, this approach is typically overlooked for breaking time reversal symmetry in photonic quantum materials due to significant single pass loss. Nonetheless, in a particular frequency band of interest, the fundamental limit on Faraday rotation compared to optical loss is favorable: for a typical Alkali metal atom like Rubidium (see appendix~\ref{App:LimitTBreak}), the ratio of intrinsic atomic linewidth to D-line fine structure is $\sim10^{-5}$, providing $\sim 10^5$ cycles of time-reversal-broken dynamics (for example, cyclotron orbits) within a photon lifetime (see appendix~\ref{App:IsoTheory}). Towards this end, early work realized small magneto-optic rotations in free-space atomic vapors \cite{franke2001magneto}.

\begin{figure}[!ht]
\includegraphics[width=8cm]{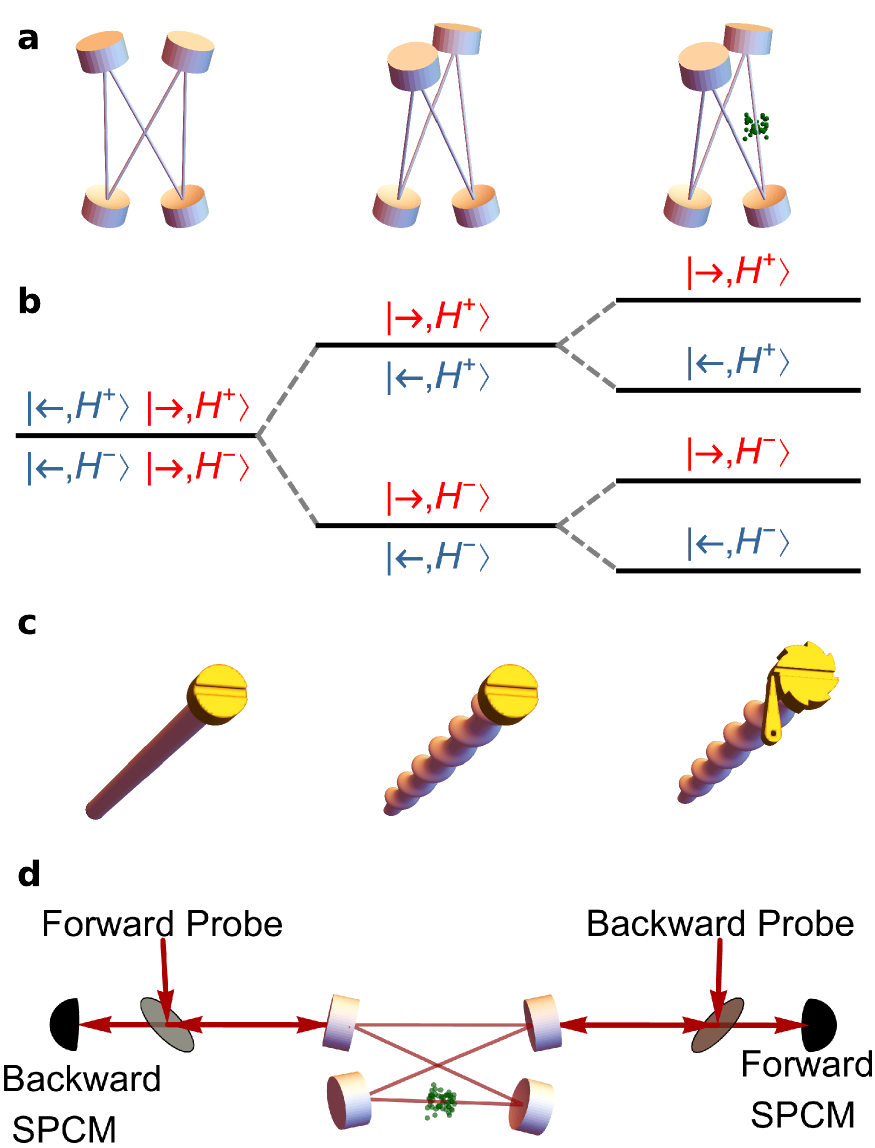}
\caption{(Color online) \textbf{T-Breaking in Twisted Resonators Coupled to Atoms}. In a birefringence-free planar resonator (\textbf{a}, left) each transverse mode exhibits a four-fold degeneracy that may be parametrized as forward (red right arrow) and backward (blue left arrow) propagation for each of positive and negative helicity (\textbf{b}, left): $\{\rightarrow,\leftarrow\}\otimes\{H^+,H^-\}$. Twisting the resonator breaks this four-fold degeneracy into two sub-manifolds of definite helicity (\textbf{a},\textbf{b} middle). We couple the optical modes to spin-polarized atoms (a, right) to break the forward-backward symmetry (\textbf{b}, right): polarized atoms are sensitive not to the light's helicity (defined relative to the direction of propagation) but to its absolute polarization (defined relative to a fixed axis); the difference in oscillator strengths for $\sigma^+$ and $\sigma^-$, for $^{87}$Rb atoms on the $|F_g=2,m_F=2\rangle\rightarrow |F_e=3'\rangle$ transition of the D2 line, is a factor of 15 \cite{steck2001rubidium}. The Zeeman splitting of the magnetic sublevels does not directly contribute to T-breaking, except insofar as it is employed to optically pump the atoms. \textbf{(c)} A schematic of the particular symmetries broken in the various aspects of the experiment, using the analogy of moving a rod into/out-of a plate. \textbf{Left}: A smooth rod can move into- or out-of- the page. \textbf{Center}: a {\it threaded} rod must twist clock-wise to move into the page, and counter-clock-wise to move out of the page. \textbf{Right}: a {\it ratcheted threaded} rod may only rotate clockwise, and thus may only move into the page. \textbf{(d)} The experimental apparatus consists of a twisted resonator coupled to an ensemble of laser-cooled $^{87}$Rb atoms (green spheres), and probed from both directions using laser fields injected through optical pickoffs (gray circles). The transmitted fields in both directions are detected through single photon counting modules (SPCMs) fiber-coupled to the light transmitted through the pickoffs.
}
\label{fig:setup}
\end{figure}

Multiple passes through the atomic ensemble may be employed to enhance the non-reciprocal polarization rotation~\cite{RomalisFaraday2011}, and indeed suggests that, in an optical cavity, the resonator geometry can be employed to control photon mass and trapping ~\cite{sommer2016engineering}, with a Faraday rotation to break time-reversal. The challenge is that the optical Faraday effect cancels in a two-mirror cavity where the forward and backward paths comprise the same mode, while in a three-mirror (running-wave) cavity the birefringence and polarization-dependent transmission of the mirrors enforce spectrally split linearly-polarized eigen-modes with vastly different finesses \cite{nagorny2003collective,klinner2006normal}. A cavity-enhanced non-reciprocity was recently demonstrated in a whispering gallery mode optical resonator \cite{sayrin2015nanophotonic}, where the cavity-birefringence was circumvented by coupling the atoms to the longitudinal component of the resonator near-field. In the present work, we extend these ideas, employing a four-mirror running-wave resonator that we twist slightly out of the plane, as in a non-planar ring oscillator ~\cite{NPRO1986}, to break inversion symmetry. An atomic ensemble provides a resonator-enhanced atomic Faraday effect that breaks time-reversal symmetry. Together, these broken symmetries result in a frequency shift between forward and backward propagating modes that we employ to demonstrate optical isolation. This is particularly exciting in light of the recent observation of photonic Landau levels in twisted optical resonators~\cite{schine2016synthetic,sommer2016engineering}; the technique demonstrated in this work would prevent interaction-induced backscattering between forward and backward propagating lowest Landau levels, paving the way to studies of Laughlin physics~\cite{umuc2014prob,somm2015quan,grusdt2013fractional} when a Rydberg admixture~\cite{pari2012obse,Jia2016CavityRydPol,jia2017strongly} induces interactions between the resonator photons. To isolate a single running-wave mode in an optical resonator, we begin by noting that even a single transverse mode of a running-wave optical resonator exhibits a four-fold degeneracy arising from the polarization-helicity degree of freedom, and the direction of propagation along the resonator axis (see Fig. \ref{fig:setup}(b)). It will thus be necessary to break \emph{two} symmetries to isolate precisely one of these modes: inversion symmetry and time-reversal symmetry.

\begin{figure*}
\includegraphics[width=7in]{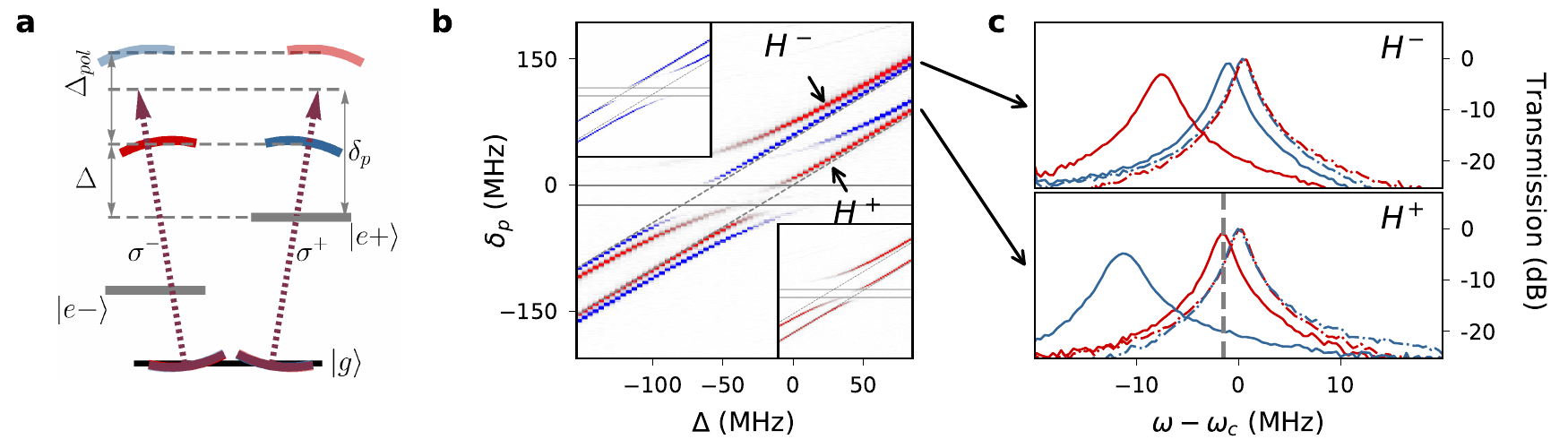}
\caption{(Color online) \textbf{Spectroscopy of a Time Reversal Broken Twisted Optical Resonator}. The relevant energy level is depicted in \textbf{(a)}. The four empty-cavity modes group into two helicity sub-manifolds separated by $\Delta_{pol}=55.5$ MHz due to the cavity twisting. The dashed purple arrows show the energy of the incident probe photons which are injected in both the forward and backward directions, and have $\sigma^-$ (left) and $\sigma^+$ (right) polarization components. In \textbf{(b)}, we show the cavity transmission (relative to the maximum empty-cavity transmission) versus the detuning $\delta_p$ of the probe from the $\sigma^+$ atomic transition and $\Delta=\Delta_{h^+,\sigma^+}$ of the $H^+$ cavity modes from the $\sigma^+$ atomic transition. With an atomic ensemble of laser cooled $^{87}$Rb atoms that are optically pumped into $\ket{5^2S_{1/2},F=2,m_F=2}$ state, the two forward (blue, top left inset) and two backward (red, bottom right inset) modes exhibit four avoided crossings as each becomes resonant with the appropriate atomic transition. In the far-detuned limit, the cavity-like modes show a frequency shift with little dissipation. The two modes within the same helicity manifold (upper two traces are $H^-$, lower two traces are $H^+$) split from one another due to the difference in atomic polarizability for $\sigma^+$ and $\sigma^-$ polarized light on the $\ket{F_g=2,m_F=2} \rightarrow \ket{F_e=3,m_F=1,3}$ transitions. Zoomed-in spectra at $\Delta=86$ MHz from the Zeeman-shifted $\ket{F_g=2,m_F=2}\rightarrow\ket{F_e=3,m_F=3}$ transition for $H^-$ (top) and $H^+$ (bottom) helicities are shown in \textbf{(c)}. The preservation of time-reversal symmetry for the empty cavity is manifest in the degeneracy of the two empty cavity modes propagating in  opposite direction (blue and red dashed lines). The forward-$\sigma^+$ (left solid peak) and backward-$\sigma^-$ (right solid peak) modes of the $H^+$ manifold are shifted from their bare-cavity frequencies by 11.3 MHz and 1.7 MHz, respectively, with an optical isolation (reverse transmission suppression at the forward resonance frequency) of 20.1(4) dB shown at the vertical dashed line. In the $H^-$ manifold, the shifts are 8.1 MHz and 1.5 MHz for backward-$\sigma^+$ (left solid peak) and forward-$\sigma^-$ (right solid peak) respectively. The absence of shoulders in the backward transmission spectra at the forward resonances (and vice- versa) indicates absence of backscattering.}
\label{fig:TRSBreak}
\end{figure*}

To break inversion symmetry we twist the resonator slightly (6$^\circ$ see appendix ~\ref{App:NPC}), resulting in a Pancharatnam polarization rotation of the intra-cavity field on each round-trip through the optical resonator (see Fig. ~\ref{fig:setup}(a)). Similarly to a Dove prism or a periscope, the polarization rotation in a non-planar resonator results from the geometric rotation of any vector when parallel transported around a non-planar closed loop \cite{sommer2016engineering}. This rotation produces a splitting of 55.5 MHz between pairs of helicity modes $H^+$ and $H^-$ which are nearly circularly polarized, with a small ellipticity arising from mirror-induced birefringence (see Fig. ~\ref{fig:setup}(b), and appendix~\ref{App:NPC} for details).

The key to breaking the remaining time-reversal symmetry is that helicity is defined with respect to the direction of light propagation, and not a fixed axis in space. Forward and backward propagating modes of the \emph{same} helicity have \emph{opposite} polarization, and thus may be split through the Faraday effect. In our experiment, this takes the form of coupling to an optically pumped atomic ensemble whose polarizability depends upon the incident light's polarization, not its helicity; the atoms are indifferent to the light's direction of propagation.

The symmetry breaking mechanisms employed by the atomic ensemble in the twisted the cavity can be understood by imagining a rod move into or out-of a hole in a plate (see Fig. \ref{fig:setup}(c)). A smooth rod respects  helical symmetry since it can slide in and out without enforcing a specific rotation direction. Both movement directions are also allowed as result of preserved time reversal symmetry. Threading breaks the helical symmetry for the movement by forcing the rod to turn in a particular direction when moving longitudinally, just as the twisting does to cavity modes. Finally, the Faraday effect suppresses the back-propagating mode with the same helicity, which is analogous to having a ratchet attached to a threaded rod to prevent backward motion. 

In our experiments, we load a cloud of $\sim$1000 $^{87}$Rb atoms into the 12$\mu$m$\times$11$\mu$m TEM$_{00}$ waist of a running wave optical resonator with a linewidth of $\kappa=2\pi\times 1.5$ MHz and finesse of $F=2500$ (see Fig. ~\ref{fig:setup}(d)). A bias field of $\sim$ 14 Gauss is then applied to the atoms along the resonator axis. The atoms are optically pumped into $\left|F_g=2, m_F=2\right\rangle$ using $\perp$ polarized light tuned to the $|F_g=2\rangle\rightarrow |F_e=2'\rangle$ transition of the $D_2$ line (see appendix ~\ref{App:OptiPump} for details on the protocol). We achieve a maximal collective cooperativity $N\eta = 4 G^2/(\kappa \Gamma) \approx 590$ and collective single-quantum Rabi frequency of $G_{\sigma^+}=73$ MHz. Here $N$ is the atom number, $\eta$ is the single quantum cooperativity of the $^{87}$Rb $5$S$_{1/2}\leftrightarrow 5$P$_{3/2}$ transition~\cite{TanjiCQED2011}, and $\Gamma$ is the $^{87}$Rb 5P$_{3/2}$ spontaneous linewidth~\cite{steck2001rubidium}.

Fig. \ref{fig:TRSBreak}(a) shows the accessible states of the system, consisting of four cavity modes and two atomic excitations, where the $\sigma^{+(-)}$ polarized cavity modes drive atoms to the $|e^{+(-)}\rangle = |F_e=3,m_F=3(1)\rangle$ levels of the $5P_{3/2}$ excited state. We write the detuning of the cavity modes with helicity $h$ from the atomic transition of polarization $\sigma$ as $\Delta_{h,\sigma}$ and define the reference cavity detuning as $\Delta=\Delta_{h^+,\sigma^+}$.

In Fig. \ref{fig:TRSBreak}(b), we experimentally explore time-reversal symmetry breaking in the cavity/probe detuning ($\Delta/\delta_p$) plane. Without the atomic ensemble, there are two pairs of degenerate empty cavity modes (diagonal dashed lines) split by $55.5$ MHz due to resonator twist. Within each pair, the forward (blue, top left inset) and a backward (red, bottom right inset) modes have the same helicity but opposite polarization relative to a fixed axis. When the atoms are transported into the cavity, they break time reversal symmetry through two independent effects: first, the two light polarizations are resonant with their respective atomic transitions at frequencies that differ by 26 MHz resulting from the differential Zeeman shift of the atomic levels of the excited state; second, the two atomic transitions have substantially different coupling strengths~\cite{steck2001rubidium}, leading to different vacuum Rabi splittings for the two polarizations. The observed spectra are in good agreement with theoretical expectations (see appendix ~\ref{App:TheorySpectrum})

In order to reduce the loss and enhance isolation (reverse transmission suppression at the forward resonance frequency), we operate the system at large detuning from $F_g=2 \rightarrow F_e=3$ transition (Fig. \ref{fig:TRSBreak}(c)). When $\Delta_{h,\sigma}$ is large compared with the collective light-matter single-excitation Rabi frequency $G_{\sigma}=\sqrt{N}g_{\sigma}$, the cavity resonances shift by $G_{\sigma}^2/\Delta_{h,\sigma}$, splitting the $\sigma^+$ and $\sigma^-$ modes for a given helicity. Here $g_{\sigma}$ is the effective single-atom vacuum Rabi coupling of the ensemble of $N$ atoms.  Splitting the forward and backward modes of a given helicity relies primarily upon the ratio $\alpha=g_{\sigma^-}/g_{\sigma^+}$ of the light-matter coupling strengths, arising from the differential polarizability of the atomic ensemble. For the states chosen above, the ratio is $\frac{1}{15}$ near the $F_g=2\rightarrow F_e=3$ transition of the D$_2$ line, and $\frac{1}{3}$ at large detunings compared to the excited state hyperfine splitting (see appendix \ref{App:LimitTBreak}). Meanwhile free-space scattering is substantially suppressed due to the large detuning $|\Delta_{h,\sigma}| \gg \Gamma$, where $\Gamma$ is the linewidth of the excited atomic state. 

\begin{figure*}
\includegraphics[width=18cm]{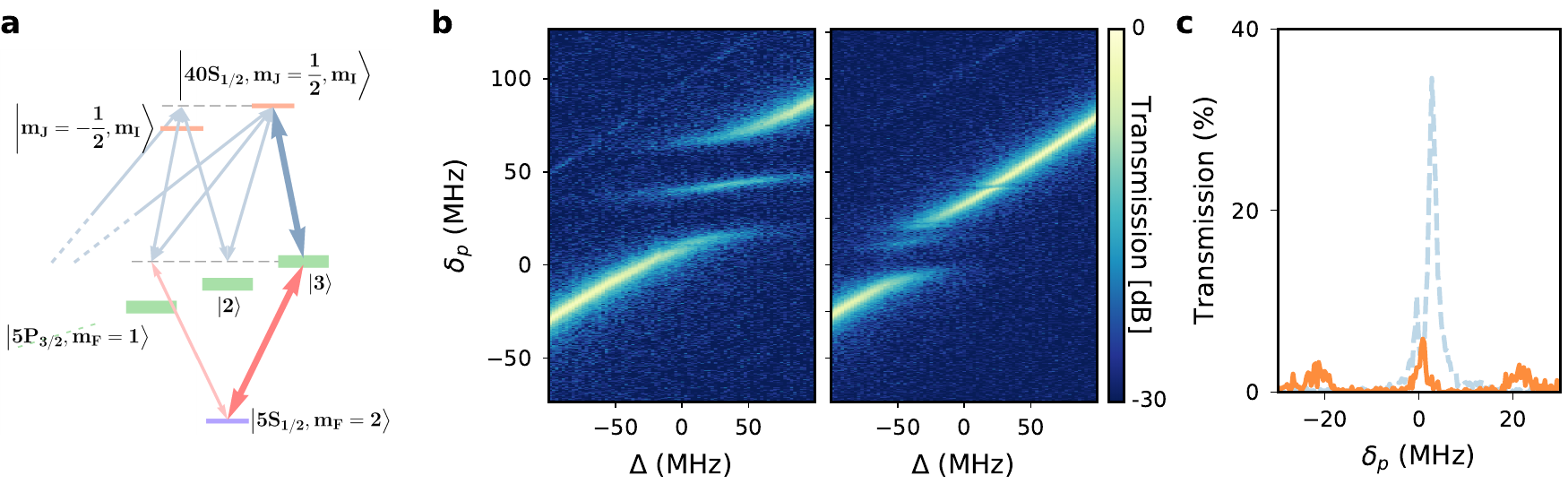}
\caption{
\textbf{Time Reversal Breaking in Cavity Rydberg EIT}. (a) The atomic level structure and relevant atomic transitions for cavity Rydberg Electromagnetically Induced Transparency. An ensemble of optically pumped $^{87}$Rb atoms are coupled to the cavity modes on the $5S_{1/2}\leftrightarrow 5P_{3/2}$ transition in the presence of a control field which further couples the atoms to the $4S$ Rydberg manifold. The small hyperfine structure $<$MHz of the Rydberg manifold places it in the Paschen-Back regime (with $m_J$ a good quantum number), while the ground- and P-state manifolds, with their larger hyperfine structure of $\sim 6.8$GHz and $\sim 500$MHz respectively~\cite{steck2001rubidium}, reside in the Zeeman regime (with $m_F$ a good quantum number). In the forward mode, both electron and nuclei spins are polarized so only stretched states couplings are allowed, and the system forms a standard 3-level EIT diagram. The backward mode is more complicated due to the non vanishing coupling between $40S,m_J=1/2\leftrightarrow 5P_{3/2}, F=3, m_F=0,1,2$ P-states with only $m_F=1$ strongly coupled to the cavity field. (b) Normalized cavity transmission vs probe ($\delta_p$) and cavity ($\Delta$) detunings. When the cavity is on resonance with the $\sigma^+$ mode, probing in the forward direction (left panel) reveals a standard EIT spectrum~\cite{Jia2016CavityRydPol}; probing backwards reveals only a weak Fano feature for this cavity tuning. The coupling of the backwards cavity field to the $m_J=-1/2$ Rydberg manifold is apparent from the prominent Zeeman shifted EIT feature in that spectrum, and the corresponding Fano feature in the forward spectrum. A slice at $\Delta=0$ is shown in (c), exhibiting vacuum Rabi splittings and an EIT feature in the forward spectrum (orange solid), and only a single Fano resonance in the reverse spectrum (blue dashed).}
\label{fig:rEITtBreak}
\end{figure*}

In Fig. \ref{fig:TRSBreak}(c), we show the transmission spectrum with the cavity 86 MHz detuned from the Zeeman-shifted $\ket{F_g=2,m_F=2}\rightarrow\ket{F_e=3,m_F=3}$ transition. Without the atoms, respect for time-reversal symmetry is manifest in the absence of any difference between forward and backwards traces (blue and red dashed lines, respectively). With the addition of an optically pumped atomic ensemble, time-reversal symmetry is broken, as shown in the solid traces, where we observe a shift of $9.4$ MHz for probing in one direction relative to the other, in agreement with $(1-\alpha^2)\frac{G^2}{\Delta}=13.8$ MHz from a simple first-principles theory. To demonstrate that the system behaves as a narrow-band optical isolator, we select a single isolated transmitting mode as the ``forward'' direction of our isolator, and measure both the reduction of transmission compared with the maximum transmission of the empty resonator, and the ``reverse" transmission of the mode in the same helicity-manifold as the ``worst-case'' isolation. We observe an isolation of 20 dB for the backward-mode and a forward-mode transmission-reduction of only $17(1)\%$; the chosen detuning deviates from the theoretical optimum derived in appendix \ref{App:IsoTheory} due to a breakdown of the approximation $G\ll\Delta$, mirror birefringence, and contributions from other excited hyperfine states. 

It is now interesting to examine the impact of backscattering and T-breaking on Rydberg polaritons, crucial ingredients of both photonic quantum information processors and quantum materials. While Rydberg-polariton collisions should be protected from backscattering by the translational symmetry of the atomic cloud, finite size and imperfect uniformity of the cloud violate this symmetry (see appendix ~\ref{App:EITBackScattering} for details), and so it is worthwhile to explore the density of backwards propagating polariton modes, along with their character.

The linear susceptibility vanishes on EIT resonance~\cite{flei2000dark}, so it is natural to anticipate that the Faraday effect that we have so fruitfully exploited for T-breaking will also vanish on EIT resonance. We find this to be true, up to T-breaking shifts from other hyperfine states; furthermore, there is no requirement that the behavior \emph{near} EIT resonance must respect timer reversal symmetry, and so residual T-breaking in EIT takes the form of different polariton properties: a change in the dark-state rotation angle and polariton loss.

The relevant transitions for the experimental investigation are depicted in Fig. \ref{fig:rEITtBreak}(a). The forward (thick) and backward (thin) modes couple to different atomic P-state magnetic sub-levels, and different Rydberg states (with different $m_I$); they \emph{must} couple to the same $m_j(=\frac{1}{2})$ in the Rydberg manifold, as the large magnetic field tunes the other $m_j(=-\frac{1}{2})$ state away by many MHz. The result is that forward mode forms a closed three-level system due to the fully polarized- electron and nuclear- spins ($|F=3,m_F=3\rangle$ in the P-manifold can only couple to $|m_J=\frac{1}{2},m_I=\frac{3}{2}\rangle$ in the Rydberg manifold), while the Rydberg state in the backward mode is coupled through the control field to other $|5P,m_F\rangle$ states which behave as loss channels and provide additional shifts and loss for the dark polariton.

Figure \ref{fig:rEITtBreak}(b) shows the experimentally observed forward (left) and backward (right) EIT spectra versus both probe frequency $\delta_p$ and resonator frequency $\Delta$; the most prominent feature is the shift of the vacuum Rabi (bright polariton) peaks due to the atomic Faraday effect, akin to the two-level case explored in Fig. ~\ref{fig:TRSBreak}(b). The dominant feature induced by the control-field coupling to the third (Rydberg) manifold is the appearance of \emph{dark} polariton resonances for $m_J=\pm\frac{1}{2}$ Rydberg states, Zeeman shifted with respect to one another; the $m_J=\frac{1}{2}$ polariton is most visible in the forward spectrum, with the $m_J=-\frac{1}{2}$ polariton clearer in the backward spectrum. Nonetheless, as anticipated, \emph{both} polaritons are visible in both spectra: as shown in Fig. \ref{fig:rEITtBreak}(c), there is a weak Fano-like feature in the backwards cavity spectrum near frequency of the $m_J=\frac{1}{2}$ dark polariton, resulting from the control field coupling between $|40S_{1/2},m_j=1/2\rangle$ and $|5P_{3/2},m_F=1\rangle$.

This Fano feature is the backward channel into which forward polaritons may scatter, and it is apparent that the backward polaritons are more weakly coupled, arising from a light-matter coupling of $G=10$ MHz (versus $G=18$ MHz in the forward direction, determined by the Clebsch-Gordon coefficients~\cite{steck2001rubidium} and reflected in the vacuum Rabi splittings of \ref{fig:rEITtBreak}b) and a $480$ nm coupling field Rabi frequency of $\Omega=4.2$ MHz (versus $\Omega=9.6$ MHz in the forward direction); this difference is visible in the width of the EIT windows,  manifested as the cavity detuning range over which the EIT and Fano features persist: $40$ MHz in the forward direction, and $10$ MHz in the backward direction.

We have demonstrated the isolation of a single time-reversal broken mode in a low-loss running-wave optical resonator. We break inversion symmetry by twisting the resonator out of the plane and time-reversal symmetry by  coupling the resonator modes to an optically active atomic ensemble. We have employed our technique to create a narrowband optical isolator with a line-width of $\sim$1.5 MHz, an isolation of 20 dB, and a relative transmission of $\sim$83\%. This performance may be straightforwardly enhanced by (1) increasing the density of atoms trapped within the resonator waist to enhance the T-breaking; (2) operating at larger detuning to reduce the loss; and (3) employing a resonator with a larger twist to enhance the inversion-symmetry breaking and reduce the impact of mirror birefringence on mode circularity. Indeed, with a higher collective cooperativity (by employing a high finesse cavity coupling to larger number of atoms), we expect transmission efficiency of $\sim$80\% and isolation of $\sim$ 60 dB (see appendix \ref{App:TheorySpectrum} for details).

Extending the technique to cavity Rydberg polaritons, the essential building block of strongly correlated cavity-based photonic materials~\cite{somm2015quan}, we see that the Faraday rotation vanishes on EIT resonance, but that the the forward-$\sigma$+ EIT window is nonetheless wider and less lossy than its time-reversed partner, which appears as a weak, shifted Fano feature; in the case of a thin atomic sample with the potential to weakly backscatter due to Rydberg-Rydberg interactions, this difference in density of states will suppress backscattering-- for additional suppression a cloud of optically pumped $^{85}$Rb atoms could be loaded into the resonator to provide a Faraday rotation while avoiding an EIT resonance due to the isotope shift. In conjunction with recently observed synthetic magnetic fields for photons \cite{schine2016synthetic} and strong interactions between individual resonator photons\cite{jia2017strongly}, this work demonstrates that all essential elements are now in place for studies of topologically ordered states of light in the fractional quantum Hall regime.

\begin{acknowledgments}
The authors thank Michael Cervia, William Tahoe Schrader, and Michelle Chalupnik for contributions to the experimental system. A.R. acknowledges support from the NDSEG fellowship. The U.S. DOE grant FP054241-01-PR supported apparatus construction, as well as data-collection and analysis; AFOSR MURI grant FP062752-01-PR supported theoretical modeling.
\end{acknowledgments}

\bibliography{polaritons}
\bibliographystyle{apsrev4-1}

\appendix
\section{Non-planar Cavity}
\label{App:NPC}
A sufficiently (compared to mirror birefringence) non-planar ring cavity exhibits circularly polarized eigenmodes arising from a round-trip geometric polarization rotation. This may be seen by parallel-transporting a polarization vector through the path described by the cavity mode and comparing with the initial vector. Numerically, this is accomplished via the round-trip transfer matrix in a 4x4 ABCD-matrix formalism \cite{schine2016synthetic}. This formalism assigns an appropriate matrix operator to propagation between- and reflection off of- each mirror. The product of matrices corresponding to travel around the cavity is the transfer matrix, and its application to a vector in the reference plane describes that vector's stroboscopic temporal evolution. The stroboscopic rotation of the vector is then gauge-independent and the result of a geometric phase.

This formalism treats polarization and image vectors differently, as none of propagation, mirror curvature, or astigmatism affect the polarization vectors; polarization vectors only reflect at each mirror. While in general calculating the round trip rotation angle is tedious, simple analytic expressions have been computed for highly symmetric geometries \cite{sommer2016engineering}.

Our experimental four-mirror cavity was designed to be planar; a small misalignment during assembly introduced a small non-planarity of the cavity mode resulting in the reported 55.5 MHz circular polarization mode splitting. Numerical modeling of our apparatus following the ABCD-matrix formalism indicates approximately $6^{\circ}$ non-planar misalignment between the upper and lower axes of the cavity. 

That a few degree misalignment can result in significant circular polarization mode splitting points to near-negligible linear birefringence of our mirror coatings, which, in subsequent fully planar cavities~\cite{jia2017strongly}, serves to split linear polarization modes by several MHz. By contrast, image astigmatism due to non-normal reflections off of curved mirrors still dominates over image rotation, resulting in very nearly pure Hermite-Gaussian transverse mode profiles. Much more significant non-planarity must be employed to overcome mirror astigmatism and provide Laguerre-Gaussian-like cavity modes, which are eigenstates of orbital angular momentum \cite{schine2016synthetic}.

\section{Performance Limit of T-Breaking}
\label{App:LimitTBreak}
For an atomic sample optically pumped into the $|5^2S_{1/2},F_g=2,m_F=2\rangle$ state, all D-line coupling to $\sigma^+$ polarized light comes from the $|5^2P_{3/2},F_e=3,m_F=3\rangle$ state, at a wavelength of ~780nm, with a (relative) Clebsch-Gordon coupling of 1. In contrast, the $\sigma^-$ coupling comes from $|5^2P_{3/2},F_e={3,2,1},m_F=1\rangle$ states at $\sim$ 780nm, and $|5^2P_{1/2},F_e={2,1},m_F=1\rangle$ states, at a $\sim$ 795nm; the sum of the squares of all of these Clebsch-Gordon's is also unity, indicating that, at very large detunings from the atomic line, there is no Faraday rotation.

That the \emph{total} atom-light coupling strengths at large detunings are equal for $\sigma^+$ and $\sigma^-$ light should come as no surprise; for light which is sufficiently detuned that it cannot couple to electron or nuclear spins through spin-orbit or hyperfine interactions, the atom appears to be a scalar scatterer and as such cannot distinguish optical polarizations. This is the same reason that state-dependent optical lattices operate most efficiently near the D-lines ~\cite{mandel2003controlled}.

\begin{figure}
\includegraphics[width=8cm]{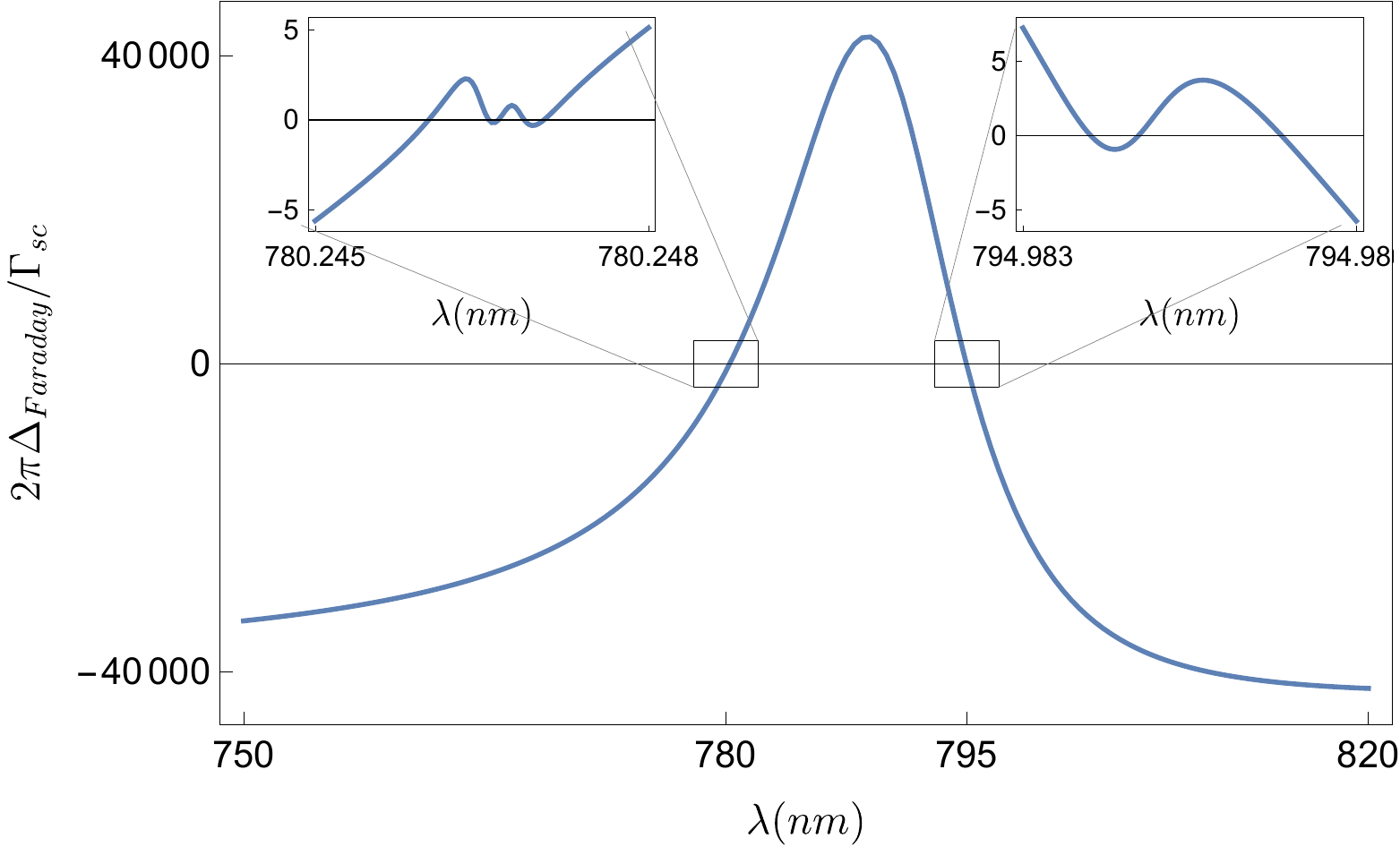}
\caption{(Color online) \textbf{Ratio of Faraday Rotation to Loss}. The competition between Faraday rotation of an optical field (which breaks time reversal symmetry) and atomic scattering of the field (which induces loss of the photons), for $^{87}$Rb atoms in the $\ket{F=2,m_F=2}$ ground state, is shown as a function of wavelength of the optical field. Away from the $D_1$ and $D_2$ atomic lines (at 795nm and 780nm, respectively), the ratio saturates, as the atomic scattering scales as $\Delta^{-2}$, and for detunings larger than the atomic fine-structure, the Faraday rotation also scales as $\Delta^{-2}$, where $\Delta$ is the approximate detuning to the atomic lines. At smaller detunings, the Faraday rotation grows as $\Delta^{-1}$, while the scattering grows as $\Delta^{-2}$, enhancing loss relative to time-reversal breaking. The zoomed-in panels show zero crossings when each excited hyperfine features is traversed.
}
\label{SI_FIG:RbFaraday}
\end{figure}

The total cavity shift for a $\sigma^+$ mode is thus $\frac{G^2}{\Delta}$, where $\Delta$ is the mode detuning from the $|5^2P_{3/2},F_e=3,m_F=3\rangle$ state. By contrast, the cavity shift for $\sigma^-$ modes is $G^2\times\sum_{F_e,J_e={1/2,3/2}}{CG(2,2,F_e,1;J=1/2,J_e)^2\frac{1}{\omega-\omega(F_e,J_e)}}$. If we assume that we are near-detuned to the $D_2$ ($J=3/2$) line  compared with the fine-structure ($\sim$ 15nm in Rubidium), the $D_1$ ($J=1/2$) Clebsch-Gordons do not contribute. If we are additionally at a large detuning compared with the excited-state hyperfine structure ($\sim$ 600MHz in $^{87}$Rb D2 line), the detuning factor $\omega-\omega(F_e,3/2)$ becomes largely independent of $F_e$, and may be written as $\Delta$; what remains is the sum of squares of Clebsch-Gordon coefficients, which in this case equal 1/3; this suggests that for optimal Faraday rotation per atom, while minimizing scattering-induced loss, one should choose a detuning which is on the order of, but not larger than, the atomic fine structure. Figure \ref{SI_FIG:RbFaraday} shows the (numerically computed) ratio of the Faraday dispersive shift to the scattering rate near each D-line feature and as well as over a broad range of detunings.

\section{Figures of Merit For Isolation}
\label{App:IsoTheory}
Consider the time reversal doublet of modes $|F,H^+\rangle$,$|B,H^-\rangle$, each with a bare linewidth $\kappa$, coupled to atoms whose spontaneous linewidth is $\Gamma$ and collective atom-light coupling strengths are $G$ and $\alpha G$ for forward and backwards modes, respectively (for a $^{87}$Rb atomic ensemble optically pumped into $F_g=2,m_F=2$, $\alpha^2=\frac{1}{3}$ for the large detunings from $D_2$ excited state hyperfine structure, and $\frac{1}{15}$ for detunings close to the $F_e=3$ hyperfine feature \cite{steck2001rubidium}). Here we assume that the coupling strength is not detuning dependent-- that is, that we are either at a small detuning compared with the hyperfine structure, or a large one, but not in-between, and similarly for the atomic fine structure. We would like to know how much the two modes may be spectrally separated by atom-induced dispersion, while maintaining at least $(1-\epsilon)$ of the empty-cavity transmission in each mode.

A detuning $\Delta$ between the cavity modes and the atomic line induces mode frequency shifts of $\frac{G^2}{\Delta}$ and $\frac{\alpha^2 G^2}{\Delta}$, neglecting the differential Zeeman shifts of the $\sigma^+$ and $\sigma^-$ atomic lines; the differential resonance shift is thus $\delta_{cav}\equiv(1-\alpha^2)\times\frac{G^2}{\Delta}=(1-\alpha^2) N\eta\frac{\Gamma}{4\Delta}\kappa$, where $N\eta\equiv \frac{4 G^2}{\kappa\Gamma}$ is the resonator-enhanced collective cooperativity.

The atom-induced broadening of each cavity mode is at most $\Gamma_{sc}\equiv\frac{G^2}{\Delta^2}\Gamma=N\eta \frac{\Gamma^2}{4\Delta^2}\kappa$ (assuming $\alpha\leq1$). The reduction in resonant cavity transmission, from the empty-cavity value, is $T=\left[\frac{\kappa}{\kappa+\Gamma_{sc}}\right]^2$, and so maintaining $T\geq1-\epsilon$ requires $\Gamma_{sc}\leq \frac{1}{2}\epsilon\kappa$.

The transmission constraint imposes a lower-limit on the atom-cavity detuning of $\Delta_{min}=\Gamma\sqrt{\frac{N\eta}{2\epsilon}}$, and thus an upper limit on the T-breaking cavity shift of $\delta_{cav}^{max}=(1-\alpha^2)\sqrt{\frac{\epsilon N\eta}{8}}\kappa$, corresponding to a suppression of backwards-mode transmission, at the frequency of the forwards-mode resonance, of approximately $\left(\frac{\kappa/2}{\delta_{cav}^{max}}\right)^2=\frac{1}{(1-\alpha^2)^2}\frac{2}{\epsilon N\eta}$.

On the other hand, if the only quantity of interest is the \emph{ratio} of forward transmission to backwards transmission, the optimum is different: under these circumstances, the differential shift, in linewidths, is $\beta=\frac{(1-\alpha^2) N\eta\frac{\Gamma}{4\Delta}\kappa}{\kappa+N\eta \frac{\Gamma^2}{4\Delta^2}\kappa}$, providing a transmission ratio on the forward resonance of $R=\frac{T_f}{T_b}=\beta^2=\left[\frac{(1-\alpha^2)N\eta\frac{\Gamma}{4\Delta}}{1+N\eta\frac{\Gamma^2}{4\Delta^2}}\right]^2$. It is apparently favorable to employ as many atoms as possible ($N\eta\rightarrow\infty$), in which case $R\rightarrow\frac{4\Delta}{\Gamma}$; thus we see that insofar as the differential cavity shift falls off only inversely with the detuning, it is favorable to go to arbitrarily large detuning. As explored in appendix \ref{App:LimitTBreak}, this scaling saturates at $\Delta=\Delta_{FS}$, the fine-structure splitting. At fixed optical depth $N\eta$, the optimal detuning is $\Delta=\sqrt{N\eta}\frac{\Gamma}{2}$, and the optimal ratio $\beta=\frac{1-\alpha^2}{4}\sqrt{N\eta}$.

For engineering a synthetic material, the quantity of importance is the number of \emph{linewidths} of time-reversal symmetry breaking, as this provides the available dynamic range for interactions and single-photon physics - any interaction or single-photon process which is \emph{larger} than the T-breaking can potentially scatter into the T-broken Hamiltonian manifold.

\begin{figure}
\includegraphics[width=7.5cm]{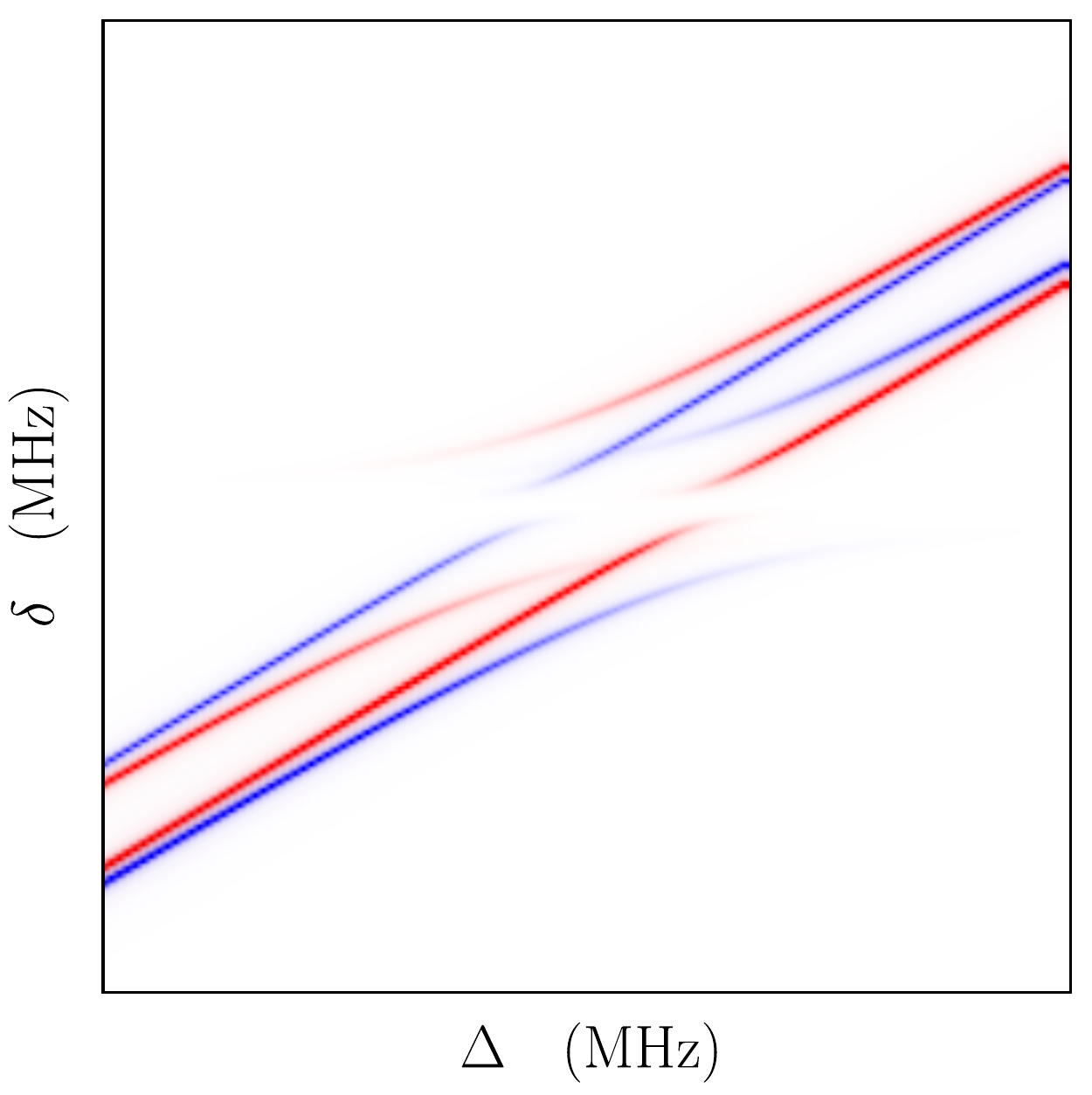}
\caption{(Color online) \textbf{Theory Spectrum of Time Reversal Breaking Cavity Transmission}. The cavity transmission of forward (blue) and backward (red) modes are shown in the cavity-detuning/probe-detuning $\Delta/\delta_p$ plane. When an atom ensemble is transported into the cavity waist, the cavity modes show four avoid crossings at distinct locations due to the Zeeman shift and cavity twist induced splitting. The difference in coupling strength between the two modes with the same helicity results in a splitting at large detuning, which is the source of the T-breaking mechanism.}
\label{SI_FIG:FullSpec}
\end{figure}

\begin{figure}
\includegraphics[width=7.5cm]{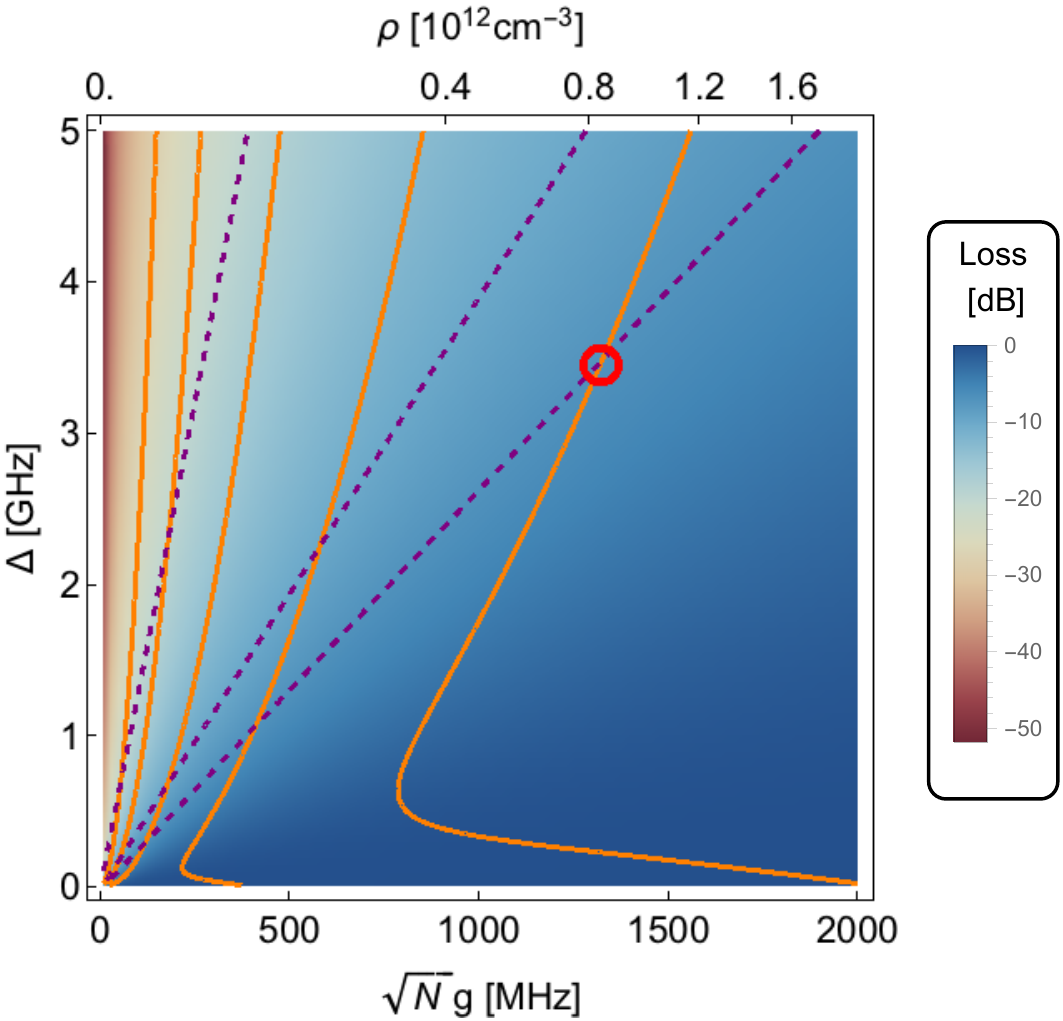}
\caption{(Color online) \textbf{Insertion Loss and Isolation}. A predicted insertion loss (density plot) and isolation (solid orange line) in the coupling strength (atomic density)/detuning plane is calculated using the non-Hermitian perturbation theory. From left to right, the isolation contour lines (solid orange) depicts the isolation from -20 dB to -60 dB in 10 dB step. The best isolation is primarily determined by the collective coupling constant ($\sqrt{N}g$), and a higher transmission, while maintaining the same isolation, could be achieved by increasing the detuning and atomic density. The three contour line (dashed purple) shows the transmission for 99\%, 90\%, and 80\% from left to right. As shown by the cross point in the circle, with coupling constant of $\sim$1.3 GHz and detuning of $\sim$3.5 GHz, one can realize a time reversal symmetry broken cavity with 20\% loss with 60 dB isolation.}
\label{SI_FIG:LossAndIsolation}
\end{figure}

\section{Theory Spectrum}
\label{App:TheorySpectrum}
We generate a model transmission spectrum by employing non-Hermitian perturbation theory \cite{cohe2004atom}, as described in \cite{somm2015quan}, treating forward (blue) and backward (red) spectra separately, and injecting elliptically polarized light composed of $\sigma^+$ and $\sigma^-$ polarizations, and detecting without polarization sensitivity. In practice, this means computing $\sigma^+$ and $\sigma^-$ traces separately (with adjustable power to match the measured data) and then adding them together. We assume atoms are in $|F_g=2,m_F=2\rangle$, and that the quanziation axis is aligned with the cavity axis, and that the cavity modes themselves are circularly polarized. The resulting theory spectrum is shown in Fig. \ref{SI_FIG:FullSpec}.

For cavity mode $\ket{\rightarrow(\leftarrow),H^+(H^-)}$, the coupling of the photon and atomic excited state is described by $2\times2$ matrices
\begin{equation}
H =
\begin{pmatrix}
\Delta+\Delta_{pol}-i\frac{\kappa}{2} & G/2 \\
G/2 & \delta_B-i\frac{\Gamma}{2}
\end{pmatrix}
\end{equation}
where $\Delta_{pol}$ is the splitting between different helicity manifolds generated by the cavity twist, $\delta_B$ is the Zeemann shift of the atomic state, and $G=\sqrt{N}g$ is the collective coupling constant. For the frequency definition in this paper, we choose the detuning to be zero at $\ket{F_g=2,m_F=2} \rightarrow \ket{F_e=3,m_F=3}$ transition, and thus $\Delta_{pol}=0$ MHz for H$^+$ and $55.5$ MHz for H$^-$ manifold. 

Using the same method, we can calculate the loss and isolation for a time-reversal-broken resonator. Here, we propose a cavity with a finesse of 100,000 and atomic sample with RMS size of 300 $\mu$m. In Fig. \ref{SI_FIG:LossAndIsolation}, we plot the loss (density plot) in the coupling strength (atomic density)/detuning plane, and overlay the contour lines (solid orange) for -20 dB to -60 dB (left to right) isolation. With a peak atomic density of $\sim 9 \times 10^{11}$ /cm$^{-3}$ and cavity detuning of $\sim$3.5 GHz, one can achieve an isolation of 60 dB and 80\% transmission (red circle in Fig. \ref{SI_FIG:LossAndIsolation}). A high transmission (99\%) and moderate isolation ($\sim$30 dB) can also be achieved with $\sqrt{N}g \approx 200$ MHz and $\Delta \approx 2.5$ GHz. 

\begin{figure}
\includegraphics[width=8cm]{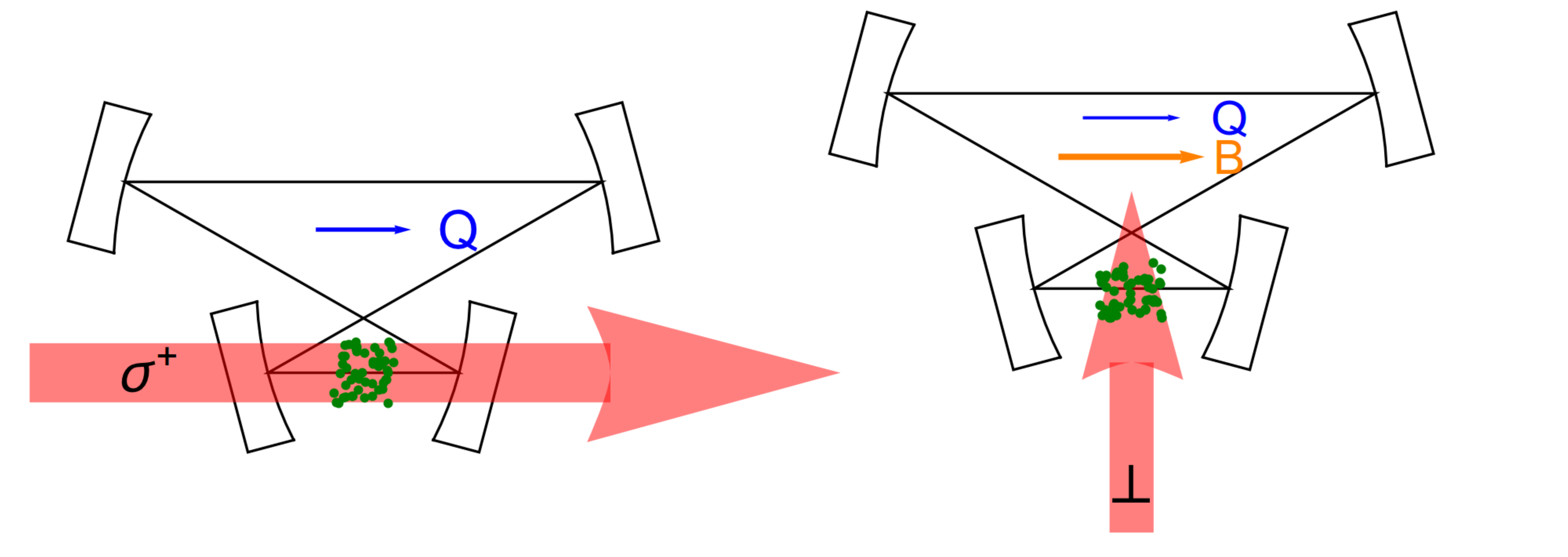}
\caption{(Color online) \textbf{Optical Pumping}. To break the degeneracy between helically polarized resonator manifolds, it is essential to achieve an atomic Faraday effect which presents a differential polarizability in the $\sigma^+$/$\sigma^-$ basis relative to a quantization axis $Q$ \emph{parallel} to the resonator axis. This requires an optical pumping beam which preferentially drives $\Delta m_F=1$ transitions, a feat which is typically achieved, as in (a), using an pumping beam which propagates propagating along $Q$, and distinguishes between $\Delta m_F=1$ and $\Delta m_F=-1$ transitions by its circular polarization. This proves difficult in our experiment, as our resonator mirrors preclude such a beam. We have developed a new technique, shown in (b), wherein the optical pumping beam distinguishes between $\Delta m_F=1$ and $\Delta m_F=-1$ transitions \emph{energetically}, via a Zeeman magnetic field $B$ parallel to $Q$, and may then propagate perpendicular to the quantization axis $Q$.}
\label{SI_FIG:OP}
\end{figure}

\section{Optical Pumping}
\label{App:OptiPump}
Our setup presents a unique optical pumping challenge: we would like to polarize our $^{87}$Rb sample in the $|F_g=2,m_F=2\rangle$ magnetic sub-level, with the quantization axis defined along the resonator axis, a task that normally require sending resonant or near-resonant circularly-polarized light through the atomic sample, with a propagation direction \emph{along the quantization axis}, as shown in Fig. \ref{SI_FIG:OP}a. The propagation direction is crucial, as we are otherwise unable to create light which polarization-selectively drives $\Delta m_F=1$ transitions. Unfortunately, we would then have to optically pump \emph{through the cavity}, which we cannot do, as the sample is optically thick in this direction, and the transverse modes of the resonator are non-degenerate. To circumvent this issue, one option is to optically pump in a rotating frame and wait for the instantaneous quantization axis to align itself with the optical resonator axis, as in reference \cite{tanji2009heralded}. For this work we have developed a new CW approach that does not rely upon a rotating spatial frame, but instead upon spectrally isolating $\Delta m_F=1$ transitions from $\Delta m_F=-1$ transitions using a Zeeman field:

We apply a magnetic bias-field along the resonator axis ($\hat{x}$) of strength 14 Gauss, and illuminate the atoms with optical pumping and re-pumping light propagating orthogonal to the resonator axis, along the transport axis ($\hat{z}$), with orthogonal linear polarization along the $\hat{y}$ axis (see Fig. \ref{SI_FIG:OP}b). The pumping light is tuned +20 MHz above the zero field $F_g=2\rightarrow F_e=2$ transition, making it resonant with the average $\delta m_F=1$ Zeeman resonance, and 40 MHz detuned from the $\delta m_F=-1$ Zeeman resonance. Atoms accumulate in the ``dark'' $|F_g=2,m_F=2\rangle$ Zeeman sub-level, with only a small scattering rate induced by the off-resonant $\delta m_F=-1$ laser field. The repumping field is tuned +17 MHz from the zero-field $F_g=1\rightarrow F_e=2$ transition frequency. 

\section{Theory of interaction-induced back scattering}
\label{App:EITBackScattering}

\subsection{Single Mode Calculation with contact interactions}
\label{App:MatrixEle}
The cavity holds two pairs of degenerate modes that are protected by the time reversal symmetry, and the polariton-polariton interaction can thus couple the forward mode to the backward mode. To investigate how this interaction induced backscattering affect the targeting many-body states without time reversal symmetry, we calculate the amplitude of the back scattering and compare it with the forward one. With a simple model of interactions, the coupling strength of forward-backward process is significantly smaller than the forward-forward one.

As long as the interaction energy is small compare to the energy difference between dark and bright polaritons, it is convenient and allowable to project the Hamiltonian onto the dark polariton manifold. In the case of a resonator with only a single transverse mode, the polariton creation operator is:
\begin{equation}
\Psi_{f,b}^\dagger =  \cos{\frac{\theta_d}{2}} a^\dagger + \sin{\frac{\theta_d}{2}} \int dz d\vec{x} \sqrt{\rho(z)} \psi(\vec{x}) r^\dagger(\vec{x},z) e^{i k_{f,b} z},
\end{equation}
where $\theta_d=\arctan{\frac{\sqrt{N}g}{\Omega}}$ is dark state rotation angle; $a$ and $r(\vec{x},z)$ represent (respectively) the annihilation operators of a cavity photon and a Rydberg excitation with transverse coordinate $\vec{x}$ and longitudinal position $z$; $\rho(z)$ is the atomic density distribution along the resonator axis; and $\psi(x)$ is the transverse mode function of the resonator. We assume that the cavity waist is much smaller than the atomic cloud, and accordingly employ a uniform atomic density distribution in the transverse plane.

The polaritons interact with each other through their Rydberg components. In the simple case of contact interactions:
\begin{equation}
V = \frac{V_{0}}{2} \int dx dz r^{\dagger}(x,z) r^{\dagger}(x,z) r(x,z) r(x,z).
\end{equation}
Since no photon operator explicitly appears in the interaction operator, it is convenient to investigate only the Rydberg-Rydberg component of the two-polariton states for the scattering calculation: $|F\rangle=(\Psi_f^\dagger)^2|\Omega\rangle$ is two forward-propagating polaritons, and $|B\rangle=(\Psi_b^\dagger)^2|\Omega\rangle$ is two backwards-propagating polaritons; $|\Omega\rangle$ is the vacuum state with no intracavity photons, and all atoms in the ground state. The amplitude for forward $\rightarrow$ backward and forward $\rightarrow$ forward scattering processes may then be computed according to:

\begin{equation}
\begin{aligned}
S_{F,B} &= \left< B \right| V \left| F \right> \\
&= 2 V_{0} \cos^2{\frac{\theta_{d_B}}{2}} \cos^2{\frac{\theta_{d_F}}{2}}
\int dz e^{i2(k_{F}-k_{B})z} \rho^2(z) \\
S_{F,F} &= \left< F \right| V \left| F \right> \\
&= 2 V_{0} \cos^4{\frac{\theta_{d_F}}{2}}
\int dz \rho^2(z)
\end{aligned}
\end{equation}
We model the atomic density distribution with a Gaussian profile:
\begin{equation}
\rho(z) = \frac{1}{\sqrt{2\pi l^2}}\exp{(-\frac{z^{2}}{2 l^{2}})}
\end{equation}
where $l$ is the length of the sample. The ratio of the forward and backward scattering amplitude is then proportional to the longitudinal phase matching term in the integral. It is straightforward to compute that the ratio of scattering amplitudes is (using $\Delta k\equiv k_f-k_b$):
\begin{equation}
\frac{S_{F,B}}{S_{F,F}} = \frac{\cos^2{\frac{\theta_{d_B}}{2}}}{\cos^2{\frac{\theta_{d_F}}{2}}}\exp(-{\Delta k}^2 l^2).
\end{equation}
In our experiment, the length of the sample is $\sim$10 $\mu$m, $k_{f,b}=k_{coupling}\pm k_c$, so $\Delta k=2 k_c=\frac{4\pi}{\lambda_c}$, and $\lambda_c=780$nm, so the backwards-to-forwards scattering ratio is $\sim$10$^{-11000}$ -- negligibly small. The real-world graininess of the atomic cloud (explored below) will impose a more physical limit on backscattering.

\subsection{Multimode Calculation with Finite Range Interactions}
We now consider a multi-mode cavity where the system behaves as massive particles in a harmonic trap ~\cite{somm2015quan}, and investigate the collision between spatially localized polaritons separated transverse to the cavity axis. Under these (non-spatially-overlapping) conditions it makes sense to consider a finite-range interaction potential between Rydberg atoms of the form $V(\vec{r},\vec{r}')=V(|\vec{r}-\vec{r}'|)$. We now write out the forward- and backward- propagating collective Rydberg creation operators localized at $\vec{r}_{2d}$ transverse to the cavity axis: $\Psi_r^{{f,b}\dagger}(\vec{r}_{2d})=\int\sqrt{\rho(z)}dz e^{i k_{f,b} z} \psi_r^\dagger(\vec{r}_{2d},z)$; here $\psi_r^\dagger(\vec{r})$ promotes an atom at location $\vec{r}$ from the ground state to the Rydberg state, $\rho(z)$ is the (normalized) distribution of the atomic density along the resonator axis, and the integral along the resonator (z)-axis reflects the fact that a polariton may be localized in the multimode cavity \emph{transverse} to the cavity axis but always remains delocalized longitudinally over the full extent of the atomic cloud propagating either forward- or backward- in the running wave cavity with wave-vector $k_{f,b}$ respectively, in keeping with the low-energy Floquet manifold in which the physics occurs~\cite{sommer2016engineering}.

The Rydberg interaction Hamiltonian may then be written: $H_{int}=\int d\vec{r} d\vec{r}' \psi_r^\dagger(\vec{r})\psi_r^\dagger(\vec{r}')V(\vec{r},\vec{r}')\psi_r(\vec{r}')\psi_r(\vec{r})$

We now consider the interaction matrix element between two forward-propagating collective Rydberg excitations at (2d coordinates) $\vec{x}_{A,B}$ and either two forward- or two backward- propagating collective excitations at (2d coordinates) $\vec{x}_{C,D}$. The collisional coupling takes the form:

\begin{equation}
\Omega_{AB\rightarrow CD}^{ff\rightarrow(ff,bb)}=\langle \Omega|\Psi_r^{f,b}(\vec{x}_C)\Psi_r^{f,b}(\vec{x}_D) | H_{int} |\Psi_r^{f\dagger}(\vec{x}_A)\Psi_r^{f\dagger}(\vec{x}_B) | \Omega \rangle
\end{equation}

where $|\Omega\rangle$ is a ``vacuum'' configuration where all atoms are in the ground state. Substituting in the definitions of the various operators yields:

\begin{widetext}
\begin{multline}
\Omega_{AB\rightarrow CD}^{ff\rightarrow(ff,bb)}=\int d\vec{r} d\vec{r}'\,dz_A\,dz_B\,dz_C\,dz_D e^{i k_f \left((z_A+z_B)\mp(z_C+z_D)\right)} \sqrt{\rho(z_A)\rho(z_B)\rho(z_C)\rho(z_D)}V(|\vec{r}-\vec{r}'|)
\\ \left[\delta(\vec{r}-\vec{x}_A)\delta(\vec{r}'-\vec{x}_B)+\delta(\vec{r}-\vec{x}_B)\delta(\vec{r}'-\vec{x}_A)\right] \left[\delta(\vec{r}-\vec{x}_C)\delta(\vec{r}'-\vec{x}_D)+\delta(\vec{r}-\vec{x}_D)\delta(\vec{r}'-\vec{x}_C)\right]
\end{multline}

where $\mp$ corresponds to forward and backward scattering, respectively. Further simplification yields:

\begin{multline}
\Omega_{AB\rightarrow CD}^{ff\rightarrow(ff,bb)}=\left[\delta(\vec{x}_A-\vec{x}_C)\delta(\vec{x}_B-\vec{x}_D)+\delta(\vec{x}_A-\vec{x}_D)\delta(\vec{x}_B-\vec{x}_C)\right] \times \left\{\begin{array}{lr}
        1 & \text{for } ff\rightarrow ff\\
        \exp(-{\Delta k}^2 l^2) & \text{for } ff\rightarrow bb
        \end{array}\right\} \times \tilde{V}(|\vec{x}_A-\vec{x}_B|)
\end{multline}
\end{widetext}

where $\tilde{V}(\Delta)$ is the effective 2D potential between collective Rydberg excitations separated by a 2D distance $\Delta$, which for $V(r)=\frac{C_6}{r^6}$ takes the form:

\begin{multline}
\tilde{V}(\Delta)=\frac{C_6}{64\sqrt{\pi}\Delta^5 l^5}[e^{\Delta^2/4l^2}\pi(\Delta^4-4\Delta^2l^2+12l^4)\erfc(\frac{\Delta}{2l})
\\-2\sqrt{\pi}\Delta l(\Delta^2-6l^2)]
\end{multline}

which behaves approximately as $\sim 1.03\times\frac{C_6}{\Delta^6}$ for large separations ($\Delta \gg l$) and $\sim 0.34\times\frac{C_6}{\Delta^5 l}$ for small separations ($\Delta \ll l$) . Independent of the form of $\tilde{V}(\Delta)$, it is apparent that $\frac{\Omega_{AB\rightarrow CD}^{ff\rightarrow(bb)}}{\Omega_{AB\rightarrow CD}^{ff\rightarrow(ff)}}=\exp(-{\Delta k}^2 l^2)$, akin to the case of contact interactions.

\subsection{Backscattering due to Atom-Cloud Graininess}
To incorporate the discreteness of the atoms into the calculation from the preceding section, we replace integrals over (coarse-grained) atomic-excitation creation operators with sums over atom locations, resulting in an interaction operator: $H_{int}=\sum_{jk}\psi_{j}^\dagger\psi_k^\dagger V_{jk}\psi_k\psi_j$, and collective Rydberg creation operator: $\Psi_r^{{f,b}\dagger}(\vec{r}_{2d})=\sum_{j\in \vec{r}_{2d}} \psi_j^\dagger e^{i k_{f,b}z_j}$; note that the last sum is only over atoms located at $\vec{r}_{2d}$, the transverse location of the polariton; we assume that there are $N_{atoms}$ such atoms. This approach circumvents the added complexity of explicitly including the transverse wave-function of the polariton in the calculation.

The discreteness of the atoms produces Rydberg-Rydberg scattering amplitude with a random phase (depending upon the particular realization of the atomic distribution). Accordingly, computing $\langle\Omega_{AB\rightarrow CD}^{ff\rightarrow bb}\rangle$ using the discrete notation, and performing an ensemble average, yields the result from the preceding section. On the other hand, the r.m.s. value of the scattering amplitude $\sqrt{\langle|\Omega_{AB\rightarrow CD}^{ff\rightarrow bb}|^2\rangle}$ reflects the discreteness of the atomic distribution. The result is (ignoring the exponentially suppressed term derived in the preceding section):

\begin{multline}
\sqrt{\langle|\Omega_{AB\rightarrow CD}^{ff\rightarrow bb}|^2\rangle} \approx \frac{1}{N_{atoms}} \tilde{V_1}(|\vec{x}_A-\vec{x}_B|)
\\ \times \left[\delta(\vec{x}_A-\vec{x}_C)\delta(\vec{x}_B-\vec{x}_D)+\delta(\vec{x}_A-\vec{x}_D)\delta(\vec{x}_B-\vec{x}_C)\right]
\end{multline}

Here $V_1(\Delta)$ behaves like $1.2\frac{C_6}{r^6}$ for large separations ($\Delta \gg l$) and $0.56\frac{C_6}{r^{5.5} l^{0.5}}$ for small separations. Accordingly:
\begin{equation}
\frac{\sqrt{\langle|\Omega_{AB\rightarrow CD}^{ff\rightarrow bb}|^2\rangle}}{\Omega_{AB\rightarrow CD}^{ff\rightarrow ff}}\approx \frac{1}{N_{atoms}}
\end{equation}

It is thus clear that graininess of the atom distribution results in collisional backscattering only once in every $~N_{atoms}^2$ collisions, where $N_{atoms}$ is the number of atoms participating in each polariton. In short, so long as the polaritons are not excessively localized, this effect will also be ignorable: even in ~\cite{jia2017strongly}, the strongly interacting cavity polaritons comprise approximately 150 atoms, so more than $10^4$ collisions would be required to produce a single backscattering event.

A similar (though simpler) calculation for two-level atoms reveals that the forward-propagating collective P-state couples to the backwards cavity mode with the single atom light-matter coupling strength $g$, as compared to its collectively-enhanced coupling to the forward cavity mode of strength $G=g\sqrt{N_{atoms}}$. Put another way, the backwards linear scattering amplitude from single atoms is smaller by a factor of $\sqrt{N_{atoms}}$, as compared to a factor of $N_{atoms}$ for the nonlinear (Rydberg-mediated collision) backscattering amplitude computed above. While the linear backscattering is only suppressed by a factor of $N_{atoms}$, as opposed to $N_{atoms}^2$ for the collisional backscattering, one backscattered polariton in every $150$ remains a relatively rare occurrence.

\subsection{Phase-Matched Linear Backscattering}
It is apparent that linear backscattering is suppressed by terms of the form $\sum_l e^{i (k_f-k_b) z_l}$, while collisional backscattering is suppressed by terms of the form  =$\sum_{lm} e^{i (k_f-k_b) (z_l-z_m)}$. In either case, if the atoms were distributed such that the summand always had the same sign, backscattering would be collectively enhanced. This process is equivalent to Bragg scattering by the atomic cloud, and as such relies upon the atoms being preferentially located on a lattice with a spacing which is a multiple of $\frac{\lambda}{2}$, as observed by ~\cite{klinner2006normal}; this scenario should be carefully avoided.

Note: In all of the above, we only consider collective states with longitudinal momenta $k_{f,b}$ because the cavity Floquet manifolds which are nearly energetically-degenerate with the EIT control fields occur only at these longitudinal momenta. There are certainly cavity modes with other longitudinal momenta, but they are at very different energies, and so do not couple resonantly to the P- and Rydberg- collective states of the same spatial structure. As a consequence, the EIT control field moves the polaritons out of resonance, and Rydberg-mediated interactions do not resonantly induce scattering into these states. This is the same mechanism that suppresses longitudinal Doppler decoherence of cavity Rydberg polaritons~\cite{Jia2016CavityRydPol}.

\end{document}